\documentstyle[12pt,colordvi,epsfig]{article}
\newcommand{\be}{\begin{eqnarray}}
\newcommand{\ee}{\end{eqnarray}}
\newcommand{\ud}{\underline}
\newcommand{\nd}{\noindent}
\begin{document}
\begin{center}
\medskip
\medskip
\medskip          
\Large{\bf $J/\psi$ Gluonic Dissociation Revisited : II.\\
Hydrodynamic Expansion Effects}
\vskip 0.2in
\large{ B. K. Patra$^1$ and V. J. Menon$^2$}
\vskip 0.2in
\normalsize{$^1$ Dept. of Physics, Indian Institute of Technology,
Roorkee 247 667, India\\
$^2$ Dept. of Physics, Banaras Hindu University, 
Varanasi 221 005, India}
\end{center}
\vskip 0.3in

\begin{center}
Abstract
\end{center}
We explicitly take into account the effect of hydrodynamic expansion
profile on the gluonic breakup of $J/\psi$'s produced in an equilibrating
parton plasma. Attention is paid to the space-time inhomogeneities as well
as Lorentz frames while deriving new expressions for the gluon number
density $n_g$, average dissociation rate $\langle \tilde{\Gamma} \rangle$,
and $\psi$ survival probability $S$. A novel type of partial
wave {\em interference} mechanism is found to operate in the formula
of $\langle \tilde{\Gamma} \rangle$. 
Nonrelativistic longitudinal expansion fro small length of the initial
cylinder is found to push the $S(p_T)$ graph above the no flow case
considered by us earlier~\cite{rev1}. However, relativistic flow corresponding
to large length of the initial cylinder pushes the curve of $S(p_T)$
downwards at LHC but upwards at RHIC. This mutually different effect
on $S(p_T)$ may be attributed to the different initial temperatures generated
at LHC and RHIC.
\vskip 0.3in

\vskip 0.3in

\newpage
\section*{I INTRODUCTION} 
Extensive literature exists on the possible suppression~\cite{review}-
\cite{langevin} of the $J/\psi$ mesons in a quark-gluon plasma and their 
proposed regeneration~\cite{Thews}.
Among the well known mechanisms of $J/\psi$ dissociation the one due to
gluonic bombardment~\cite{xu} deserves special attention here. Recently the
present authors~\cite{rev1} considered the statistical mechanics of important
physical observables {\em viz.} the gluon number density, thermally-averaged
$g-\psi$ break-up rate, and the $\psi$ meson survival probability
appropriate to RHIC/LHC initial conditions. We found~\cite{rev1} that these
observables are {\it significantly} affected if one employs improved 
expressions for 
the gluon distribution function, $g-\psi$ relative flux, and $\psi$ meson
formation time.

Of course, it is a well-recognized fact that the longitudinal/transverse
expansion of the medium controls the master rate equations~\cite{biro}
for the time-evolution of the plasma temperature and parton fugacities. But 
the literature
{\it does not} tell how the fluid velocity profile itself influences the
Lorentz transformations connecting the rest frames of the fireball, plasma,
and $\psi$ meson. In other words, since the {\it flow velocity profile} causes 
inhomogeneities in space-time, hence the scenario
of $J/\psi$ gluonic break-up may be affected in a quite nontrivial manner
and the aim of the present paper is to address this hitherto unsolved
problem.

Sec.II below recalls a few aspects of relativistic hydrodynamics for the sake 
of ready reference. Next, a new expression for the gluon number density
is derived in Sec.III showing how an extra dilation factor $\gamma$ associated 
with the flow appears. Next, in Sec.IV careful Lorentz transformations are
used to calculate the flux-weighted cross section and explicit dependence
is brought out on the hydrodynamic velocity $\vec{w}$ observed in $J/\psi$
rest frame. Next, Sec.V develops the machinery for computing the $J/\psi$
survival probability as a function of transverse momentum. Finally, 
Sec.VI summarizes our main conclusions applicable
to nonrelativistic/relativistic flows.

\section*{II ASPECTS OF HYDRODYNAMICS}

\noindent {\underline{\bf Preliminaries}}:~
Consider the rest frame of the hot, dense fireball produced in 
ultrarelativistic heavy ion collision. Within an initial time span
$t_i=\tau_0 \sim 0.5~{\rm{frm/c}}$ it is supposed to achieve local
thermal equilibrium. The plasma now expands rapidly, gets cooled at the
expense of internal energy, and is driven towards chemical 
equilibration through partonic reactions. The plasma's life ends, i.e.,
freeze-out occurs at the instant $t_{\rm{life}}$ when the temperature drops
to $200~{\rm{MeV}}$, say. Such a picture of collective flow is known
to have profound effect on the measured dilepton spectrum~\cite{munshi},
dependence of colour screening mechanism on equation of state~\cite{dipali},
anisotropy in the transverse-momentum distribution~\cite{aniso} of
output hadrons, azimuthal asymmetry of $J/\psi$ suppression in noncentral
heavy-ion collisions~\cite{azimuth}, etc.

\noindent {\underline{\bf Equation of motion}}:~
We employ the units $\hbar =c=1$ and follow closely the hydrodynamic
summary given by Pal et al~\cite{dipali} based upon {\it cylindrical symmetry} 
appropriate
to central collisions. In the fireball frame a general time-space point
$x$ and the fluid $4$ velocity $u$ have the form
\be
x=(t,\vec{x})~;~u=(\gamma, \gamma \vec{v})\nonumber\\
\gamma={(1+{\vec{u}}^2)}^{1/2}={(1-{\vec{v}}^2)}^{-1/2}
\ee
where $\vec{v}$ is the local $3$ velocity and $\gamma$ is the corresponding
Lorentz factor. Ignoring viscosity the conservation law for the energy-
momentum tensor $T^{\mu \nu}$ reads
\be
\partial_{\mu}T^{\mu \nu}=0~;~ T^{\mu \nu} (x) =(\epsilon +P) u^\mu u^\nu +
P g^{\mu \nu}
\ee
where the energy density $\epsilon$ and pressure $P$ are supposed to be 
measured in a frame comoving with the plasma. The relationship between
{\it the fireball usual time $t$ and medium proper time $\tau$} is, of course 
\be
\frac{d \tau}{dt} = \frac{1}{\gamma}~;~ t_i\le t \le t_{\rm{life}}
\ee

\noindent {\ud{\bf Longitudinal Expansion}}:~ 
In Bjorken's boost-invariant, one-dimensional flow the profile admits
a simple {\it analytical} solution
\be
\vec{v}=\frac{z}{t} \hat{e}_z~;~\tau ={(t^2-z^2)}^{1/2} \ge \tau_0
\ee
where $\hat{e}_z$ is a unit vector along the collision axis. The 
corresponding temperature $T \propto \tau^{-1/3}$ is known to
fall rather slowly, {\em e.g.} in the case of gold nuclei colliding at
RHIC. Hence the $J/\psi$ suppression remains comparatively high 
{\em via} the colour screening/gluonic dissociation mechanisms.

\noindent {\ud{\bf Transverse Expansion}}:~
As summarized in Refs.~\cite{dipali}-\cite{munshi} the $4$-velocity profile 
becomes
quite intricate and a numerical integration of the dynamical equations
(2,3) becomes very hard in the case of $3+1$ dimensional expansion. For our
purpose it will suffice to assume that the collective flow occurs only
along the lateral directions without rotation as given by the {\it empirical}
ansatz
\be
\vec{v} =\frac{r}{t} \hat{e}_r~;~ \tau ={(t^2-r^2)}^{1/2}
\ee
where $\hat{e}_r$ is the unit vector along $\vec{r}$ in a 
cylindrical coordinate system $(r,z,\phi)$. It is known  from numerical
solution of (2) that transverse flow causes very rapid cooling in the case of
lead nuclei colliding at LHC. Hence the $J/\psi$ survival chance becomes
relatively high {\em via} the colour screening/gluonic dissociation 
scenarios. We
now proceed to formulate the statistical mechanics of some physical
observables following closely the logic of Ref.\cite{rev1}.

\section*{III GLUON NUMBER DENSITY} 
\noindent {\underline{\bf Preliminaries}}:~
Working in the fireball rest frame and assuming local thermal equilibrium
let the symbol $x$ denote a typical time-space point, $u$ the medium
$4$- velocity, $T$ the absolute temperature, $K=(K^0,\vec{K})$ the gluon
$4$-momentum, 16 the spin-colour degeneracy factor, $\lambda_g$ the gluon
fugacity, $f$ the one-body Bose-Einstein distribution function, and $n_g$ 
the gluon
number density given mathematically by
\be
f=\frac{\lambda_g}{e^{K\cdot u/T}-\lambda_g}~;~ n_g(x) = 16 \int
\frac{d^3 \vec{K}}{{(2\pi)}^3} f
\ee
It may be noted that generally $\lambda_g <1$ for chemically unequilibrated
plasma and $K\cdot u$ is tedious in the fireball frame.

\noindent {\ud{\bf Comoving Frame}}:~
If $k =(k^0,\vec{k})$ is the gluon $4$-momentum in the local 
rest frame of the plasma then by Lorentz transformations
\be
K\cdot u = k^0~;~ K^0/k^0 = \gamma (1 +\vec{v}\cdot \hat{k})
\ee
where $\gamma$, $\vec{v}$ refer to the fluid motion of (1) and $\hat{k}$
is a unit vector along $\vec{k}$. Substitution into (6) simplifies
the distribution function as
\be
f=\frac{\lambda_g}{e^{k^0/T}-\lambda_g} 
= \sum_{n=1}^\infty \lambda_g^n~e^{-n k^0/T}
\ee
Also, upon taking the polar axis for integration along $\hat{v}$ and writing
$\vec{v}\cdot \hat{k} = |\vec{v}| \cos \theta_{kv}$ our number density
reduces to 
\be
n_g(x) &=& 16 \int \frac{d^3k}{{(2\pi)}^3} \frac{K^0}{k^0} f \\
&=&\frac{2}{\pi^3} \int_0^\infty dk^0 {k^0}^2 \int_{-1}^1 d \cos \theta_{kv}
\int_0^{2\pi} d \phi_{kv}~
\gamma \left( 1 + |\vec{v}| \cos \theta_{kv} \right) 
\sum_{n=1}^\infty \lambda_g^n e^{-nk^0/T} \\
&=& \frac{16}{\pi^2} \gamma T^3 \sum_{n=1}^\infty \frac{\lambda_g^n}{n^3}
\ee
\noindent \ud{\bf Remarks}:~ This result is {\it new} and shows in a compact 
manner how the
number density depends upon $\gamma$, $T$, and $\lambda_g$ (although these
dependences are generally interwoven through the evolution equations). It
may be stressed that inspite of the occurrence of a $\vec{v} \cdot \hat{k}$
term in (7) our final $n_g$ depends on $\vec{v}^2$ through $\gamma$. Our
(11) generalizes an earlier expression obtained by Xu et al~\cite{xu} who
ignored the flow velocity $\vec{v}$ completely and worked with a 
factorized form of the Bose-Einstein distribution containing $\lambda_g$
only in the numerator.

\noindent {\ud{\bf Numerical Estimate at $t_i$}}:~
The behaviour of (11) can be easily studied at the instant when the thermalized
fireball was formed in high energy heavy ion collision. The initial conditions
predicted by HIJING Monte Carlo simulations are summarized in Table 1. 

%%%%%%%%%%%% begin tab,1%%%%%%%%%%%%%%%%%%%%%%%%%%%%%%%%%%%%%%%
\begin{table}[h]
\caption{Initial values for the time, temperature,
           fugacities etc. at RHIC(1), LHC(1) only~\protect\cite{hijing}}
%\vskip 0.2in
\begin{tabular}{lcccccc}
\hline
& $T$(GeV) & $t_i=\tau_0$ (fm) & $\lambda_g$ & $\lambda_q$ &
$n_g^{v=0.1c}{\rm{(fm)}}^{-3}$ & $n_g^{v=0.9c}{\rm{(fm)}}^{-3}$ \\
   &  & & & & &\\
\hline
RHIC(1)  & 0.55 & 0.70 & 0.05 & 0.008 & 1.78 & 4.05\\
  &  & & & & &\\
LHC(1) & 0.82 & 0.5 & 0.124 & 0.02 & 14.73 & 33.63 \\
  &  & & & & &\\
\end{tabular}
\end{table}

There
the gluon densities computed {\em via} (11) in the nonrelativistic 
$(v\approx 0.1c)$
and ultrarelativistic $(v\approx 0.9c)$ regions are also listed showing a 
relative enhancement by the factor 
\be
\frac{n_{gi}^{v=0.9c}}{n_{gi}^{v=0.1c}} \approx 
\frac{{(1-0.01)}^{1/2}}{{(1-0.81)}^{1/2}} \approx 2.28
\ee

\noindent {\ud{\bf Temporal Evolution}}:~
It is very tedious to employ the exact distribution function $f$ of
(6) for determining how the fugacities and temperature evolve with the
proper time $\tau$ of the medium. Hence we shall directly borrow the 
master rate equations from existing literature based upon approximately
{\it factorized} form of $f$. It is plausible to assume that
such an approximation will not markedly affect our final conclusions
since $\lambda_g < 1$. For longitudinal expansion parametrized by (4)
the relevant ordinary differential equations~\cite{biro} are known to be
\be
&&\frac{\dot{\lambda_g}}{\lambda_g} + 3 \frac{\dot{T}}{T} +\frac{1}{\tau}
=R_3 \left(1-\lambda_g\right) - 2 R_2 \left(1-\frac{\lambda_g^2}{\lambda_q^2}
\right),\nonumber\\
&&\frac{\dot{\lambda_q}}{\lambda_q} + 3 \frac{\dot{T}}{T} +\frac{1}{\tau}
=R_2 \frac{a_1}{b_1}\left(\frac{\lambda_g}{\lambda_q} -
\frac{\lambda_q}{\lambda_g}\right),\nonumber\\
&&{\left(\lambda_g+\frac{b_2}{a_2} \lambda_q\right)}^{3/4} T^3 \tau = {\rm{\mbox{const}}}
\ee
Here $\lambda_q(\lambda_g)$ is the quark (gluon) fugacity,
$N_f$ the number of flavours, and remaining symbols are defined by
\be
&&R_2=0.5 n_g \langle v \sigma_{gg \longrightarrow q \bar{q}} \rangle, \quad
R_3=0.5 n_g \langle v \sigma_{gg \longrightarrow gg q} \rangle \nonumber\\
&&a_1=16 \zeta(3)/\pi^2,\quad  a_2=8\pi^2/15 \nonumber\\
&&b_1=9\zeta(3)N_f/\pi^2,\quad b_2=7\pi^2N_f/20
\ee

Next, for transverse expansion parametrized by (5) the appropriate
partial differential equations~\cite{dipali} read 
\be
\partial_\tau T^{00}+r^{-1}\partial_r(rT^{01})+\tau^{-1}(T^{00}+P)=0
\ee
and
\be
\partial_\tau T^{01}+r^{-1}\partial_r\left[r(T^{00}+P)v_r^2\right]+
   \tau^{-1}T^{01}
+\partial_r P=0
\ee
where
\be
T^{00}=(\epsilon+P)u^0u^0-P
\ee

Their solutions on the computer yield the functions $T(x)$,
$\lambda_g(x)$ and hence $n_g(x)$ subject to the stated
initial conditions.

\section*{IV THERMALLY-AVERAGED RATE } 
\noindent {\underline{\bf Preliminaries}}:~
Next, we turn to the question of applying statistical mechanics to
gluonic break-up of the $J/\psi$ moving inside an expanding parton
plasma. In the fireball frame consider a $\psi$ meson of mass
$m_\psi$, four momentum $p_\psi = (p_\psi^0, \vec{p}_\psi)$, three
velocity $\vec{v}_\psi$ and dilation factor $\gamma_\psi$ defined by
\be
\vec{v}_\psi = \vec{p}_\psi/p_\psi^0~,~ \gamma_\psi = p_\psi^0/m_\psi
= {\left( 1 - {\vec{v}_\psi}^2 \right)}^{-1/2}
\ee
The invariant quantum mechanical dissociation rate
$\Gamma$ for $g-\psi$ collision may be written compactly as
\be
\Gamma = v_{{}_{\rm{rel}}} \sigma 
\ee
where $v_{\rm{rel}}$ is the relative flux and $\sigma$ the cross section
measured in any chosen frame. Its  thermal average over gluon
momentum in the {\it fireball} frame reads
\be
\langle \Gamma (x) \rangle = \frac{16}{n_g(x)} \widetilde{\int}
\frac{d^3K}{{(2\pi)}^3}~\Gamma f
\ee
Here the tilde implies that the gluon is hard enough to
break $J/\psi$ and the gluonic distribution function $f$ is
evaluated at the {\it location} of the $\psi$ meson {\em viz.}
at the time-space point
\be
x=(t,\vec{x}_\psi)
\ee
The highly nontrivial integral (20) is best handled in the $\psi$ 
meson rest frame by generating several useful kinematic 
informations as follows.

\noindent {\underline{\bf Kinematics in $J/\psi$ Rest Frame}}:~
Let $q=(q^0,\vec{q})$ be the gluon $4$ momentum measured in $\psi$
meson {\it rest} frame. Since the relative flux becomes
$v_{\rm{rel}}^{\rm{Rest}} = c= 1$ hence our invariant $\Gamma$
reduces to the QCD~\cite{BP} based cross section
\be
\Gamma = \sigma_{{}_{\rm{Rest}}} = B {(Q^0-1)}^{3/2}/{Q^0}^5~;~ 
q^0 \ge \epsilon_\psi \nonumber\\
Q^0 =\frac{q^0}{\epsilon_\psi} \ge 1~;~ B=\frac{2\pi}{3}{\left(\frac{32}{3}
\right)}^2
\frac{1}{m_c {(\epsilon_\psi m_c)}^{1/2}}
\ee
here $\epsilon_\psi$ is the $J/\psi$ binding energy and $m_c$ the
charmed quark mass. This $\sigma_{{}_{\rm{Rest}}}$ possesses a sharp
peak at the gluon energy
\be
q^0_p=\frac{10 \epsilon_\psi}{7}=0.92~{\rm{GeV}}; \quad Q^0_p=\frac{10}{7}
\ee 
Also, the energy variable for the massless gluon transforms {\em via}
\be
K^0 = \gamma_\psi \left( q^0 +\vec{v}_\psi \cdot \vec{q}~\right)
= \gamma_\psi q^0 \left( 1 + |\vec{v}_\psi| \cos \theta_{q \psi} \right)
\ee
with $\theta_{q \psi}$ being the angle between $\hat{q}$ and $\hat{v}_\psi$
unit vectors. Furthermore, the fluid $4$- velocity $w = (w^0,\vec{w})$ seen
in $\psi$ rest frame will be explicitly given by the Lorentz transformations
\be
w^0 &=& \gamma_\psi \left( u^0 - \vec{u} \cdot \vec{v}_\psi \right)
= \gamma_\psi \gamma \left( 1 - \vec{v} \cdot \vec{v}_\psi 
\right) \\
\vec{w} &=& \left[ \frac{}{} \vec{u} - ( \vec{u} \cdot \hat{v}_\psi ) 
\hat{v}_\psi \frac{}{} \right] + \gamma_\psi \left[ (\vec{u} \cdot 
\hat{v}_\psi ) - u^0 |\vec{v}_\psi| \right] \hat{v}_\psi \nonumber\\
&=& \gamma \left[ \vec{v}  - \gamma_\psi \vec{v}_\psi + (\gamma_\psi -1)
(\vec{v} \cdot \hat{v}_\psi) \hat{v}_\psi \right]
\ee
Finally, the scalar product
\be
K\cdot u = q \cdot w = q^0 w^0 - q^0 |\vec{w}| \cos \theta_{qw}
\ee
where $\theta_{qw}$ is the angle between $\hat{q}$ and $\hat{w}$. We thus
have all the ingredients needed to calculate the thermally-averaged rate
of (20).

\noindent {\underline{\bf Evaluation of $\langle \Gamma(x) \rangle$}}:~
Upon taking the polar axis for $\vec{q}$ integration along $\hat{v}_\psi$,
denoting the solid angle element by $d \Omega_{q\psi}$, and expanding
$f$ in a power series we can write
\be
\langle \Gamma(x) \rangle = \frac{16}{n_g} \int \frac{d^3q}{{(2\pi)}^3}
\frac{K^0}{q^0} \sigma_{\rm{Rest}} \sum_{n=1}^\infty
\lambda_g^n e^{-n K\cdot u/T} \\
=\frac{2}{\pi^3 n_g} \int_{\epsilon_\psi}^\infty dq^0 {q^0}^2
\int_0^{4\pi} d \Omega_{q\psi} \gamma_\psi 
\left( 1+ |\vec{v}_\psi| \cos \theta_{q\psi} \right)
\sigma_{\rm{Rest}} \sum_{n=1}^\infty \lambda_g^n e^{-nq\cdot w /T}
\ee
This integral is performed in the Appendix  based on the following
nomenclature.

\noindent {\ud{\bf Symbols/Notations}}:~
Using the nomenclature in the fireball frame we define 
\be
(\theta_v, \phi_v)={\rm{polar~angles~of~}} \hat{v}~;~
({\theta}_{v_\psi}, {\phi}_{v_\psi}) = {\rm{polar~angles~of~}} \hat{v}_\psi
\nonumber\\
F = \vec{v}\cdot \hat{v}_\psi = |\vec{v}| \left[
\sin \theta_v \sin \theta_{v_\psi} \cos (\phi_v - \phi_{v_\psi} )
+ \cos \theta_v \cos \theta_{v_\psi} \right] \nonumber\\
Y = \gamma_\psi |\vec{v}_\psi| - (\gamma_\psi -1) F \nonumber\\
w^0 = \gamma \gamma_\psi (1-F |\vec{v}_\psi| )~;~
\vec{w} = \gamma (\vec{v} - Y \hat{v}_\psi) \nonumber\\
|\vec{w}|= {[{|\vec{w}|}^2]}^{1/2}= \gamma {\left[ {|\vec{v}|}^2 +Y^2 - 2YF
\right]}^{1/2} \nonumber\\
\theta_{\psi w} ={\rm{angle~between~}} \hat{v}_\psi~{\rm{and}}~ \hat{w}
\nonumber\\
\cos \theta_{\psi w} = \frac{1}{|\vec{w}|} (\vec{w} \cdot \hat{v}_\psi)
= \frac{1}{|\vec{w}|} (F-Y)
\ee
Also will be needed
\be
Q^0 = q^0/\epsilon_\psi~;~ C_n = n \epsilon_\psi w^0/T \nonumber\\
D_n = n \epsilon_\psi |\vec{w}|/T~;~ \rho_n = D_n Q^0 \nonumber\\
A_n^\pm = C_n \pm D_n = n \epsilon_\psi ( w^0 \pm |\vec{w}| ) /T
\nonumber\\
I_0 (\rho_n) = \sinh (\rho_n)/\rho_n \nonumber\\
I_1(\rho_n) = \cosh (\rho_n) /\rho_n - \sinh (\rho_n)/\rho_n^2
\nonumber\\
n K \cdot u /T = n q\cdot w/T = C_n Q^0 - \rho_n \cos \theta_{q w}
\ee

\noindent {\ud{\bf Result of Integration}}:~
From Appendix (A4) we find
\be
\langle \Gamma(x) \rangle = \frac{8 \epsilon_\psi^3 \gamma_\psi}{\pi^2 n_g} 
\sum_{n=1}^\infty \lambda_g^n \int_1^\infty
dQ^0 {Q^0}^2 \sigma_{\rm{Rest}} (Q^0) e^{-C_n Q^0} \nonumber\\
\times \left[ \frac{}{} I_0 (\rho_n) + I_1 (\rho_n) |\vec{v}_\psi| \cos \theta_{\psi w}
\frac{}{} \right]
\ee
By decomposing the hyperbolics $I_0$ and $I_1$ into exponentials this
result can also be expressed as 
\be
\langle \Gamma(x) \rangle = \frac{4 \epsilon_\psi^2 T \gamma_\psi
}{\pi^2 | \vec{w}| n_g} \sum_{n=1}^\infty \frac{\lambda_g^n}{n} 
\int_1^\infty dQ^0 Q^0 \sigma_{\rm{Rest}} (Q^0) \nonumber\\
\times \left[ \left( \frac{}{}1 + \vec{v}_\psi \cdot \hat{w} ( 1 -1/\rho_n) 
\frac{}{} \right) e^{-A_n^- Q^0} \right.\nonumber\\
\left. - \left( \frac{}{}1 - \vec{v}_\psi \cdot \hat{w} ( 1 +1/\rho_n) 
\frac{}{} \right) e^{-A_n^+ Q^0}  \right]
\ee

\noindent {\ud{\bf Remarks}}:~
The expressions (32, 33) are original and demonstrate how the mean 
dissociation rate $\langle \Gamma(x) \rangle$ depends on the
{\it hydrodynamic}  flow through $|\vec{w}|$ (or $w^0$) as well as
the angle $\theta_{\psi w}$ in the notation of (30). The structure
of $\exp(\pm D_n Q^0)$ tells that numerical treatment
of (32) is convenient if $0 \le |\vec{w}| \le T/\epsilon_\psi$ while
that of (33) is suitable if $T/\epsilon_\psi \le |\vec{w}| \le \infty$.
From the analytical viewpoint it is much more advisable to work with
the modified rate
\be
\langle \tilde{\Gamma} (x) \rangle \equiv n_g(x) \langle \Gamma(x) \rangle
\ee
This is because $\langle \tilde{\Gamma} \rangle$ is devoid of any $n_g$
factor appearing in the denominator of (32). Furthermore, 
$\langle \tilde{\Gamma} \rangle$ will be seen to enter directly the 
survival chance $e^{-W}$ in (51) later.

\nd \ud{\bf Analytical estimate}~
The physical interpretation of our graphs below will be facilitated by 
deriving a rough approximation for $\langle \tilde{\Gamma} \rangle$ 
as follows. In Eq.(32) we retain only the $n=1$ term of the summation,
and approximate the integrand by its peak value at $Q^0_p$ (cf.23) so that
the desired estimate of (34) becomes 
\be
{\langle \tilde{\Gamma}(x) \rangle} &\approx &
\frac{8 \epsilon_\psi^3 \gamma_\psi
}{\pi^2}~\lambda_g \int_1^\infty
dQ^0 {Q^0}^2 \sigma_{{}_{\rm{Rest}}}~H \nonumber\\
&\propto& \lambda_g \gamma_\psi H 
\ee
Here the {\em entire} dependence on the flow velocity $w$ is contained in the
function
\be 
H \equiv e^{-C_1 Q^0_p} \left[ \frac{}{} I_0(D_1 Q^0_p)
+I_1 (D_1 Q^0_p) \mid \vec{v}_\psi \mid \cos \theta_{\psi w} 
\frac{}{} \right]
\ee
with the coefficients $C_1$, $D_1$  and $A_1^\pm$ read-off from (31). We
are now ready to discuss some consequences of (35) in {\em three} cases
{\em viz.} static medium in the fireball frame, no flow in the
$J/\psi$ rest frame, and ultrarelativistic flow in either frame.

\noindent {\ud{\bf Static Medium in Fireball Frame}}:~
Remembering the notations (30, 31) consider the {\em hypothetical} case 
\be
\vec{v}=\vec{0}~;~ \gamma=1~;~F=0~;~ Y= \gamma_\psi |\vec{v}_\psi|
\nonumber\\
w^0 = \gamma_\psi~;~\vec{w} = -\gamma_\psi \vec{v}_\psi~;~
C_n= n \epsilon_\psi \gamma_\psi/T,~\cos \theta_{\psi w}=-1 \nonumber\\
D_n = C_n |\vec{v}_\psi|~;~ A_n^\pm = C_n (1 \pm |\vec{v}_\psi|)
\ee
This is precisely the case treated in our earlier paper~\cite{rev1}
where the $g-\psi$ relative flux was correctly taken in the $\psi$ meson
rest frame. Eqs.(32, 33) also improve the work of Xu et al~\cite{xu}
who had written the $g-\psi$ relative flux in the fireball frame. Figs.1, 2
%%%%%%%%%%%%%%%%%%% begin of figure 1 %%%%%%%%%%%%%%%%%%%%%%
\begin{figure}[h]
\psfig{file=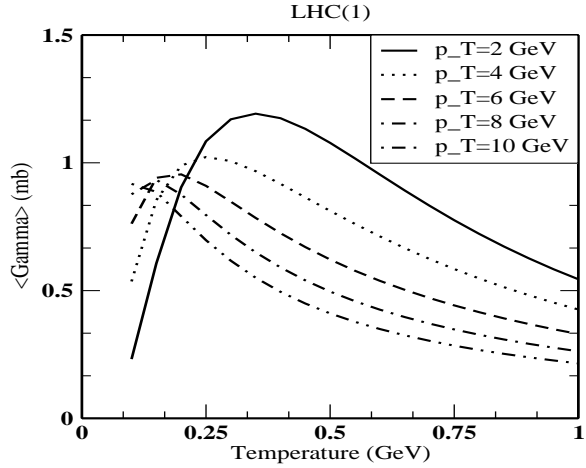,angle=0,height=12cm,width=12cm}
\caption{The thermal-averaged gluon-$\psi$ dissociation
cross section $\langle \Gamma \rangle \equiv \langle v_{{}_{\rm{rel}}} \sigma 
\rangle$ as a function of temperature at different transverse momenta
$p_T$ as in~\cite[Eq.(19)]{rev1}, i.e., in the
absence of longitudinal flow.  The initial gluon fugacity is given in Table 1 
at LHC energy.}
\end{figure}
%%%%%%%%%%%%%%%%%%% end of figure 1%%%%%%%%%%%%%%%%%%%%%%%%%
%%%%%%%%%%%%%%%%%%% begin of figure 2%%%%%%%%%%%%%%%%%%%%%%%%%
borrowed from~Ref.\cite{rev1} display the temperature and transverse momentum
\begin{figure}[h]
\psfig{file=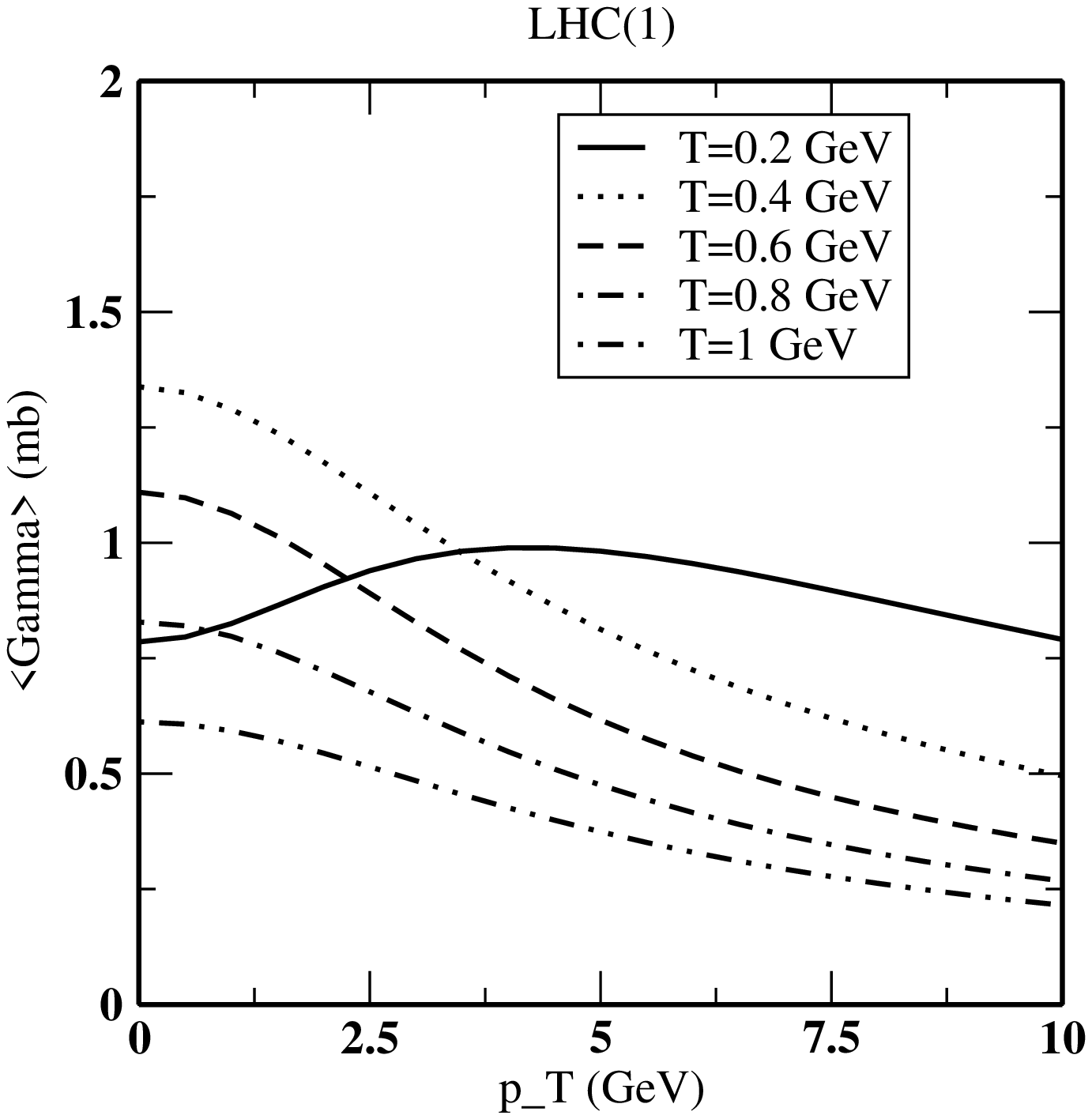,angle=0,height=12cm,width=12cm}
\caption{The variation of $\langle \Gamma \rangle \equiv \langle 
v_{{}_{\rm{rel}}} \sigma \rangle$ with transverse momentum at different
temperatures~\cite{rev1} as in Fig.1 in absence of flow profile.}
\end{figure}
%%%%%%%%%%%%%%%%%%% end of figure 2%%%%%%%%%%%%%%%%%%%%%%%%%
dependence of the {\em usual} rate $\langle \Gamma \rangle$ based on (32, 33)
relevant to the LHC(1) initial fugacity $\lambda_{gi}$. For 
the sake of comparison the corresponding curves of the {\em modified} 
rate $\langle \tilde{\Gamma} \rangle$ using (34) are drawn in Figs.3, 4.
Due to the assumed absence of flow there is no inhomogeneity with respect
to $x$ and the $J/\psi$ formation time is also ignored at this stage.
Various features of the indicated diagrams are interpreted in the next
paragraph.

%%%%%%%%%%%%%%%%%%% begin of figure 3%%%%%%%%%%%%%%%%%%%%%%%%%
\begin{figure}[h]
\psfig{file=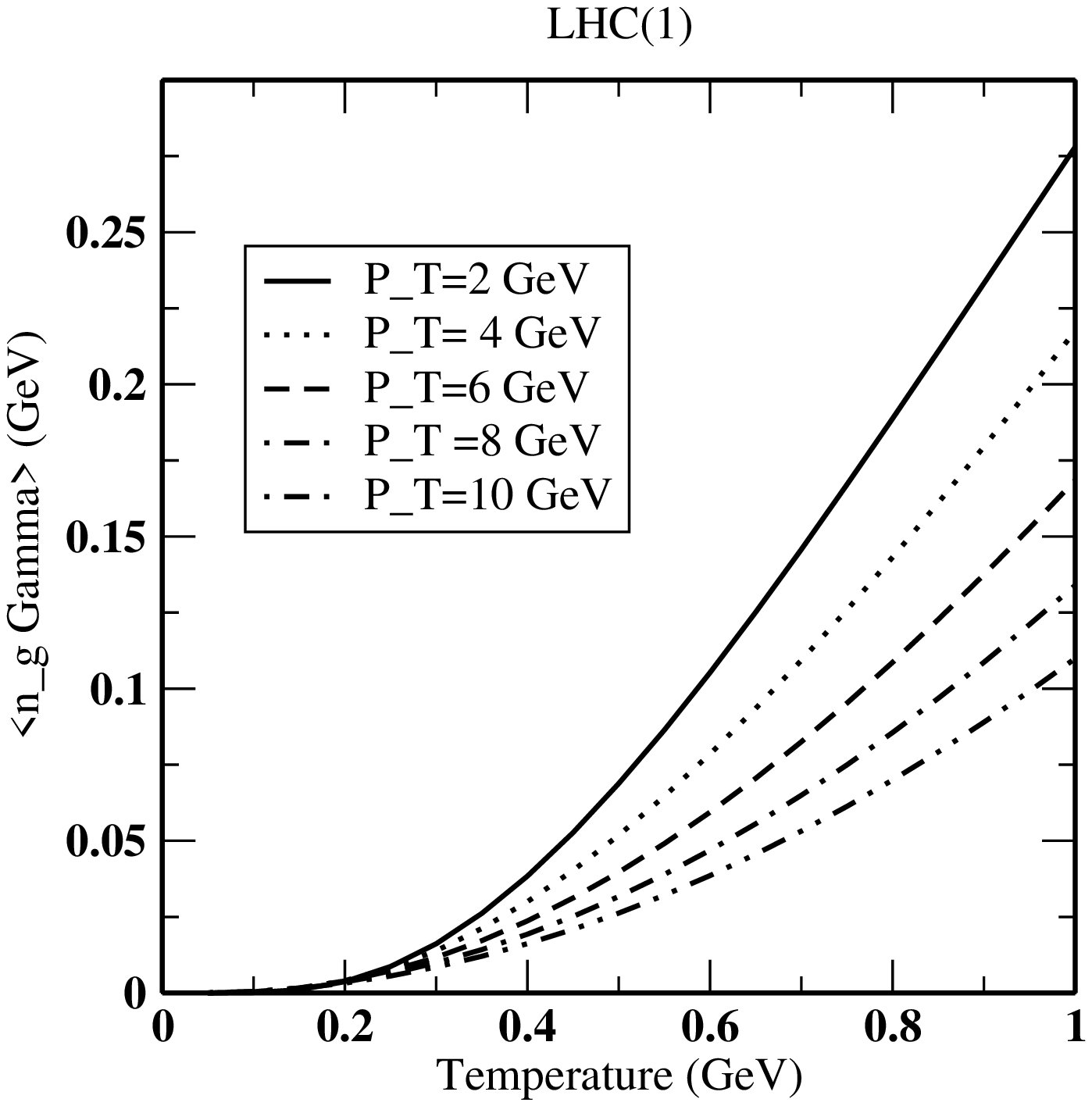,angle=0,height=12cm,width=12cm}
\caption{The variation of the {\em modified} rate
$\langle \tilde \Gamma \rangle \equiv n_g \langle \Gamma \rangle$
with temperature at different values of transverse momentum corresponding 
to $\langle \Gamma \rangle$ of Fig.1 is depicted here.}
\end{figure}
%%%%%%%%%%%%%%%%%%% end of figure 3%%%%%%%%%%%%%%%%%%%%%%%%%
%%%%%%%%%%%%%%%%%%% begin of figure 4%%%%%%%%%%%%%%%%%%%%%%%%%
\begin{figure}[h]
\psfig{file=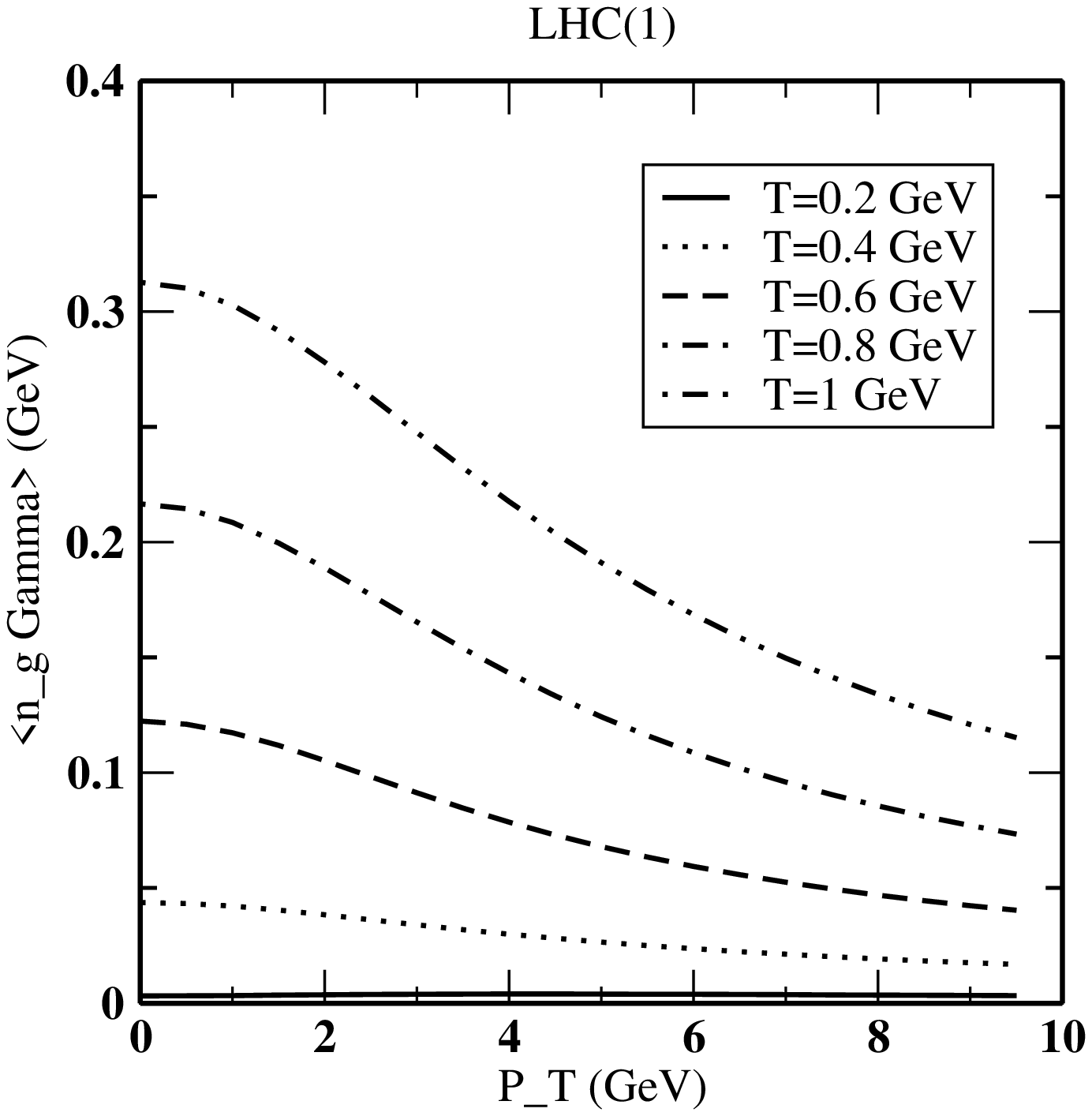,angle=0,height=12cm,width=12cm}
\caption{The variation of the {\em modified} rate
$\langle \tilde \Gamma \rangle \equiv n_g \langle \Gamma \rangle$
with transverse momentum at different values of temperatures
corresponding to $\langle \Gamma \rangle $ of Fig.2 is shown here
in absence of flow.}
\end{figure}
%%%%%%%%%%%%%%%%%%% end of figure 4%%%%%%%%%%%%%%%%%%%%%%%%%
\nd \ud{\bf Interpretation}~i) Following the arguments given by Xu et al
~\cite{xu} it is known that the peak of $\sigma_{{}_{\rm Rest}}$ at
$Q^0_p$ gives a rich structure to the usual rate $\langle \Gamma \rangle$
in Figs 1, 2. ii) However, the modified rate $\langle \tilde{\Gamma} \rangle$ 
in Figs.3, 4 becomes structureless , i.e., monotonic. This is because
the gluon number density $n_g$ in (11) contains a crucial $T^3$ factor
which changes from a low value {\em viz.} 0.008 at $T=0.2~{\rm{GeV}}$ to
a high value {\em viz.} $1.0$ at $T=1~{\rm{GeV}}$. Such a $T^3$ coefficient
matters a lot in the conversion of $\langle \Gamma \rangle$ to
$\langle \tilde{\Gamma} \rangle$  {\em via} (34). iii) At fixed $p_T$
the steady {\em increase} of $\langle \tilde{\Gamma} \rangle$ with $T$ in
Fig. 3 is caused by the growing $\exp{(-A_1^\pm Q^0_p)}$ factors of the
estimate (35, 36). iv) At fixed $T$ the monotonic {\em decrease} of   
$\langle \tilde{\Gamma} \rangle$ with $p_T$ in Fig. 4 has a very
interesting explanation. For the situation (37) under study $\vec{w} =
-\gamma_\psi \vec{v}_\psi$ is antiparallel to $\vec{v}_\psi$ so that
$\cos \theta_{\psi w} =-1$. Hence partial wave terms $I_0$ and $I_1$ of
$H$ (36) {\em interfere} destructively in Fig. 4 making 
$\langle \tilde{\Gamma} \rangle$  small as $\mid \vec{v}_\psi \mid$
grows.
 
\noindent {\ud{\bf No  Flow in $J/\psi$ Rest Frame}}:~
Next, consider a configuration in the fireball frame such that the
$3$ velocities of the plasma and $\psi$ meson coincide at some $x$.
This is possible if, for example, both the fluid and $\psi$ are
moving in the transverse direction. Then in (30, 31) we put
\be
\vec{v} = \vec{v}_\psi~;~ F=|\vec{v}_\psi|~;~ w^0=1~;~
\vec{w} = \vec{0} \nonumber\\
C_n =n \epsilon_\psi/T~;~\rho_n=0~;~I_0(\rho_n) =1~;~I_1(\rho_n) =0
\ee
Inserting this information into the exact formulae (32, 34) and attaching
a suffix $0$ we find 
\be
{\langle \tilde{\Gamma}(x) \rangle}_0= 
\frac{8 \epsilon_\psi^3 \gamma_\psi
}{\pi^2} \sum_{n=1}^\infty \lambda_g^n ~ e^{-n \epsilon_\psi/T}
\int_1^\infty dQ^0~{Q^0}^2 \sigma_{{}_{\rm{Rest}}}
\ee
Its variations with $T$ and $p_T$ are displayed in Figs.5, 6
relevant to the LHC(1) initial fugacity. The crude proportionality
%%%%%%%%%%%%%%%%%%% begin of figure 5%%%%%%%%%%%%%%%%%%%%%%%%%
\begin{figure}[h]
\psfig{file=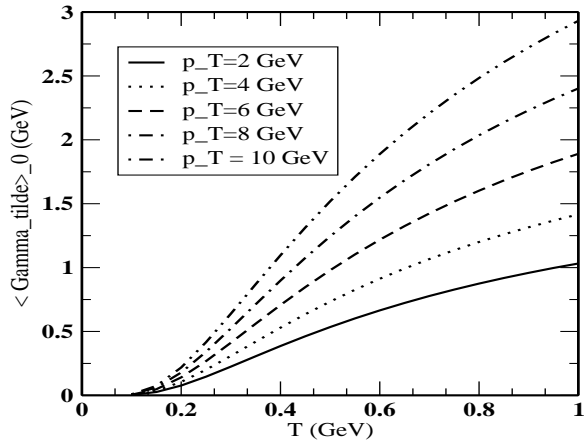,angle=0,height=12cm,width=12cm}
\caption{The variation of the {\em modified} rate  ${\langle 
\tilde{\Gamma}(x)\rangle}_0$ from Eq.(39) with temperature 
for different $p_T$'s, 
when the $3$ fluid velocity $\vec{w}$ in the $\psi$ meson rest frame
is zero.}
\end{figure}
%%%%%%%%%%%%%%%%%%% end of figure 5%%%%%%%%%%%%%%%%%%%%%%%%%
%%%%%%%%%%%%%%%%%%% begin of figure 6%%%%%%%%%%%%%%%%%%%%%%%%%
\begin{figure}[h]
\psfig{file=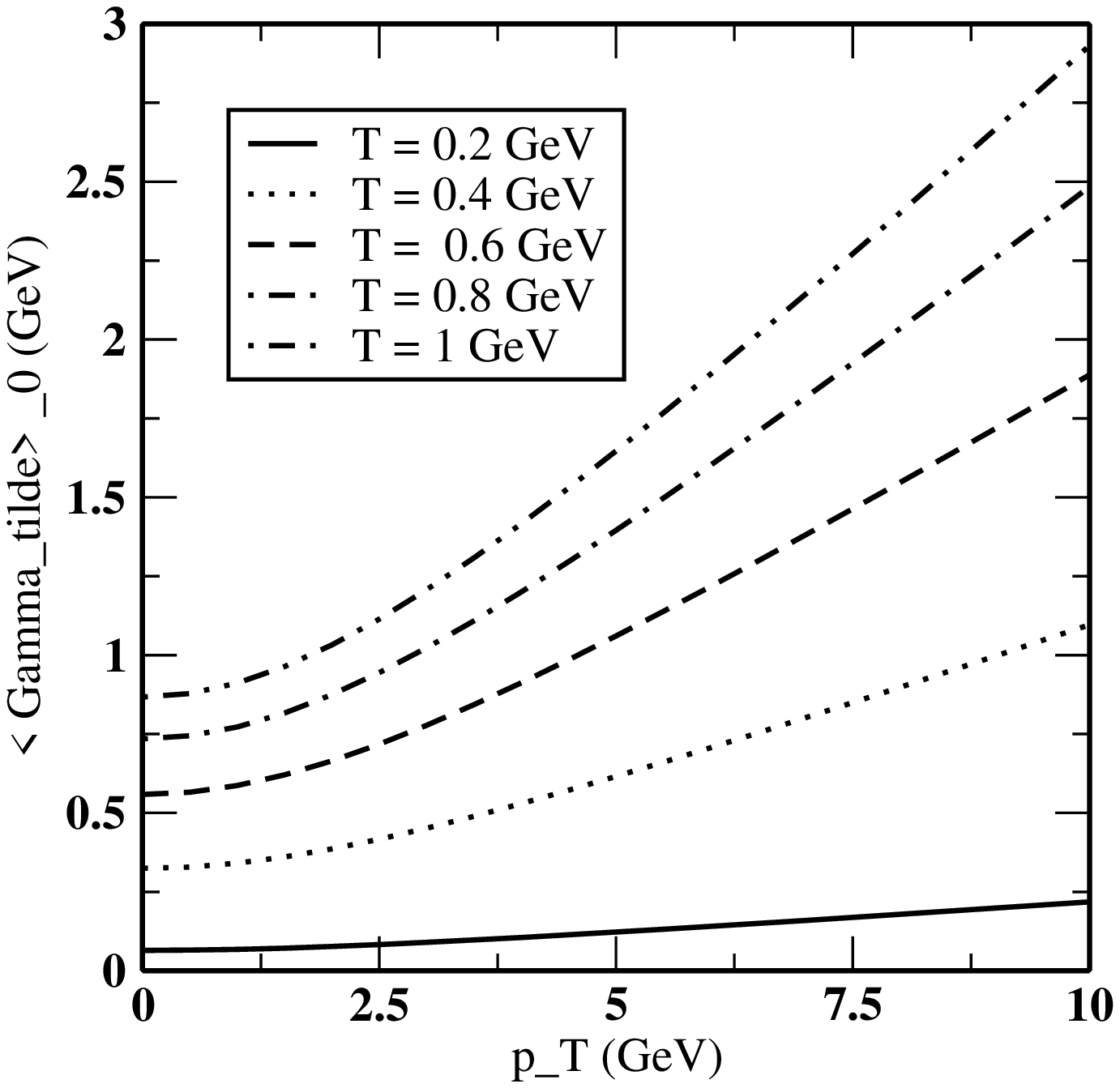,angle=0,height=12cm,width=12cm}
\caption{ The variation of the {\em modified} rate  ${\langle 
\tilde{\Gamma}(x)\rangle}_0$ from Eq. (39) with $p_T$'s for
different values of $T$'s  
when the $3$ fluid velocity $\vec{w}$ in the $\psi$ meson rest frame
is zero.}
\end{figure}
%%%%%%%%%%%%%%%%%%% end of figure 6%%%%%%%%%%%%%%%%%%%%%%%%%
(35) now reduces to 
\be
{\langle \tilde{\Gamma}(x) \rangle}_0 \propto \lambda_g \gamma_\psi 
e^{-\epsilon_\psi/T}
\ee
which is utilized below to explain some features of the graphs.

\nd \ud{\bf Interpretation}~
i) At fixed $p_T$ our ${\langle \tilde{\Gamma}(x) \rangle}_0$ 
of Fig.5 rises monotonically with $T$ in analogy with the earlier
Fig.3. This is caused by the $e^{-\epsilon_\psi/T}$ factor present
in the estimate (40). ii) At fixed $T$ our ${\langle \tilde{\Gamma}(x) 
\rangle}_0$ of Fig.6 grows steadily with $p_T$ in contrast to the
earlier Fig.4. Such a behaviour is due to the coefficient $\gamma_\psi$
occurring in (40). iii) The curves of Figs.5, 6 are consistently {\em higher}
than those of Figs.3, 4. The reason is that there is no $I_1(\rho_n)$
term present in (39) to interfere destructively with the $I_0(\rho_n)$
term in view of the restriction (38).

\noindent {\ud{\bf Ultrarelativistic Flow in Either Frame}}:~
Finally, suppose at a point $x$ in the fireball frame the medium is
flowing ultrarelativistically $(\mid\vec{v} \mid \rightarrow 1)$ and the
$J/\psi$ is moving slowly ($|\vec{v}_\psi| <1/10$, say). In
the $\psi$ meson rest frame the Lorentz transformation (26) shows that 
$\vec{w}$ almost equals $\gamma \vec{v}$ so that $\cos \theta_{\psi w}
\rightarrow \hat{v}_\psi \cdot \hat{v}$. [Hence, in the case of pure
transverse expansion of the plasma $\cos \theta_{\psi w}$ can even
become $+1$ implying a constructive interference between the $I_0$
and $I_1$ terms of (36); this possibility will, however, be discussed
in a future communication. At present, we illustrate the case of 
both the $J/\psi$ and plasma moving ultrarelativistically (in the transverse
and longitudinal directions, respectively) subject to the following
kinematic conditions:
\be
\vec{v}=0.9~\hat{e}_z;~\gamma=2.3;~ F=0;~Y=\gamma_\psi |\vec{v}_\psi| 
\nonumber\\
w^0=\gamma \gamma_\psi >>1;~~|\vec{w}| \approx \gamma \gamma_\psi -
\frac{1}{2 \gamma \gamma_\psi} \nonumber\\
\cos \theta_{\psi w} = -\frac{Y}{|\vec{w}|} \approx -\frac{1}{\gamma};
~~D_1 =\epsilon_\psi |\vec{w}|/T \sim \frac{ \epsilon_\psi \gamma 
\gamma_\psi}{T}~>>1 \nonumber\\
A_1^- =\frac{\epsilon_\psi}{T} (w^0 -|\vec{w}|) \approx \frac{\epsilon_\psi}
{2T \gamma \gamma_\psi}
\ee
Then $I_o (\rho_1) \rightarrow I_1(\rho_1) \rightarrow \exp(\rho_1)/2\rho_1$
so that our rough estimate (35, 36) become
\be
\langle \tilde{\Gamma}(x) \rangle  \propto \frac{\lambda_g T}{\gamma}
\exp \left( -\frac{\epsilon_\psi Q^0_p}{2T \gamma \gamma_\psi} \right)
\left[ 1 - \frac{{|\vec{v}_\psi|}^2}{\gamma} \right]
\ee
This information will be utilized below for explaining the main features of the
graphs.

%%%%%%%%%%%%%%%%%%% begin of figure 7%%%%%%%%%%%%%%%%%%%%%%%%%
\begin{figure}[h]
\psfig{file=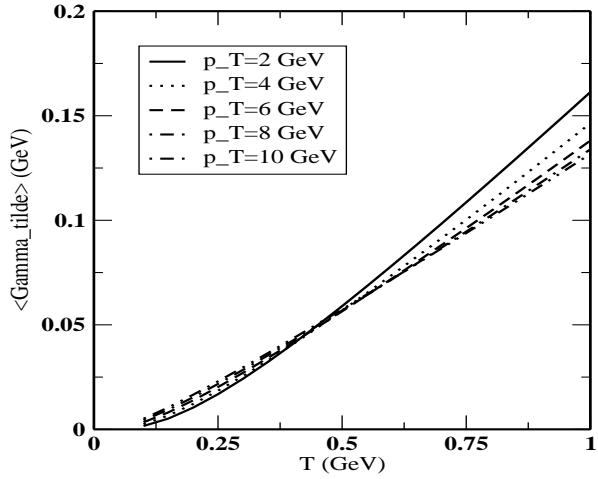,angle=0,height=12cm,width=12cm}
\caption{The variation of the {\em modified} rate  
$\langle \tilde{\Gamma}(x) \rangle$ using Eqs.(32, 34)
as a function of temperature at different transverse momenta
for the ultrarelativistic longitudinal flow velocity $v=0.9~c$.
}
\end{figure}
%%%%%%%%%%%%%%%%%%% end of figure 7%%%%%%%%%%%%%%%%%%%%%%%%%
%%%%%%%%%%%%%%%%%%% begin of figure 8%%%%%%%%%%%%%%%%%%%%%%%%%
\begin{figure}[h]
\psfig{file=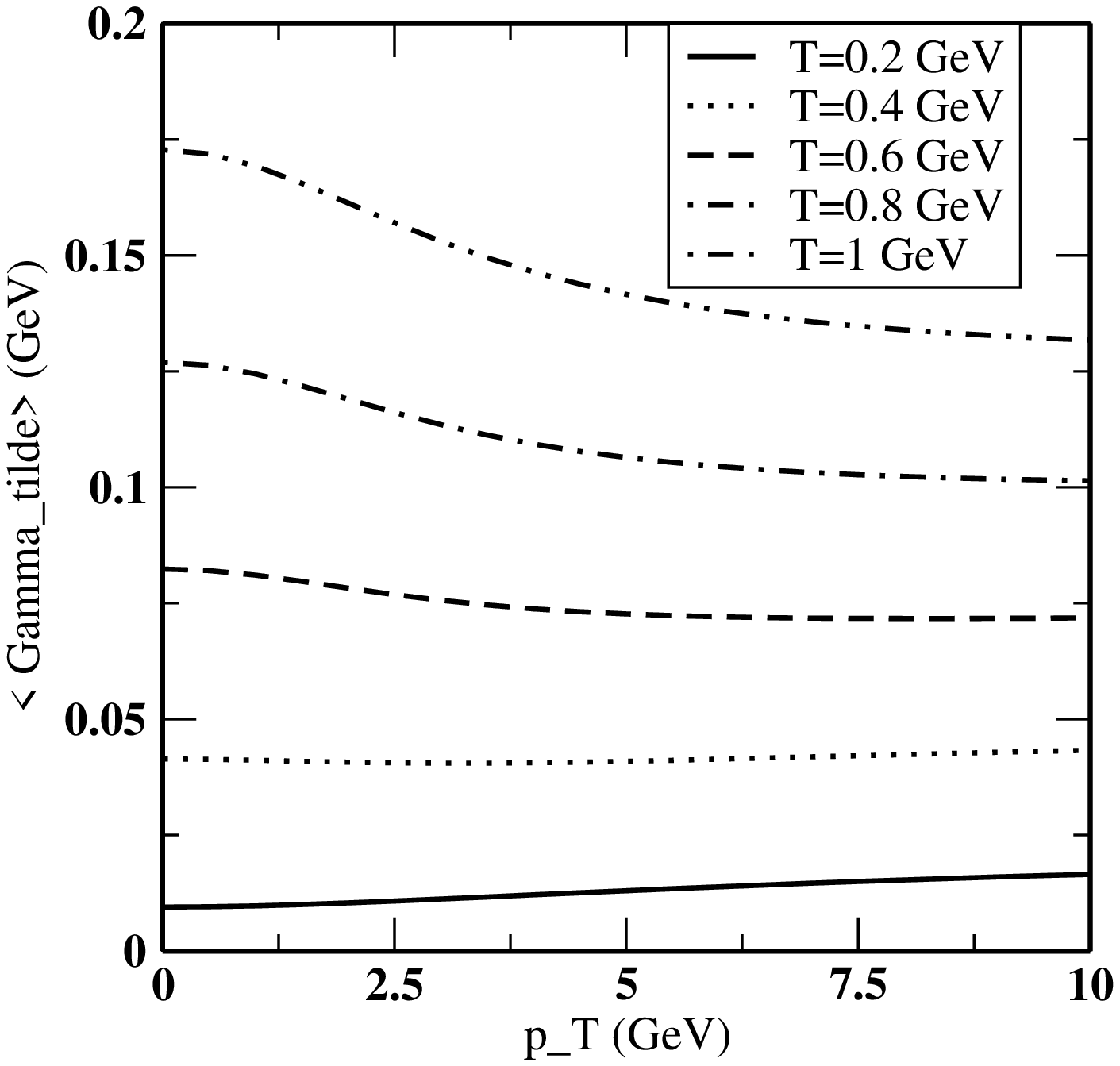,angle=0,height=12cm,width=12cm}
\caption{The variation of the {\em modified} rate  
$\langle \tilde{\Gamma}(x) \rangle$ using Eqs.(32, 34)
as a function of $p_T$ at different values of temperatures
as in Fig.7 corresponding to ultrarelativistic longitudinal flow
velocity $v=0.9~c$.}
\end{figure}
%%%%%%%%%%%%%%%%%%% end of figure 8%%%%%%%%%%%%%%%%%%%%%%%%%

\nd \ud{\bf Interpretation}
i) For fixed $p_T$, $v$ the exponential in (42) tend to $0$ as
$T \rightarrow 0$ and tends to $1$ as $T \rightarrow \infty$. 
Therefore, the growing trend
of $\langle \tilde{\Gamma}(x) \rangle$ with $T$ in Fig. 7 is understable.
ii) At fixed $T$, $v$ the rich behaviour of 
$\langle \tilde{\Gamma}(x) \rangle$ with $p_T$ in Fig. 8 arises from a
sensitive competition between the bracketed factors of (42).
iii) More precisely, at lower temperatures $T,le0.4$ GeV the exponential
factor in (42) increases dominantly with $p_T$ causing
$\langle \tilde{\Gamma}(x) \rangle$ to grow.
iv) But at higher temperatures $T \ge0.8 $ geV the third bracket in
(42) decreases prominantly with $p_T$ causing 
$\langle \tilde{\Gamma} \rangle$ to drop.
v) Comparison of Figs.(3,4) with Figs.(7,8) tells that the
$p_T$- dependence of $\langle \tilde{\Gamma} \rangle$ 
(unlike its $T$ dependence) is quite sensitive to the 
nonrelativistic/ultrarelativistic nature of flow in the fireball frame. 

\section*{V $J/\psi$ SURVIVAL PROBABILITY}

Next, we ask the important question as to how the explicit collective flow
profile affects the birth and death scenarios of $\psi$ mesons within a 
quark gluon plasma? Below we consider only {\it longitudinal} expansion; the 
case of transverse flow will be dealt with in a future communication.

\noindent {\ud{\bf Pure Longitudinal Expansion}}:~
Suppose at general instant $t$ in the 
fireball frame the plasma is contained inside a {\it cylinder} of radius $R$,
cross section $A=\pi R^2$, length $L$, and volume $V=AL$. Keeping the origin
at its centre and assuming constant speed of longitudinal expansion we have
\be
R=R_i,~A=A_i,~L=L_i t/t_i,~ V=V_it/t_i
\ee
where the suffix $i$ lebels {\it initial} values. The ansatz (4) of the 
velocity profile reads
\be
\vec{v} = z\hat{e}_z/t~;~ -L/2 \le z \le +L/2
\ee
subject to the restriction that the speed at the edges $\pm L/2$ must be less
than $c$, i.e.,
\be
\frac{L}{2t} = \frac{L_i}{2t_i} < 1
\ee

\noindent {\ud{\bf Production configuration of $J/\psi$}}:~
Employing cylindrical coordinates
let a typical $\psi$ meson be created at the instant $t_I$, having position
${\vec{x}_\psi}^I$, and with transverse velocity ${\vec{v}_\psi}^I$
(in the mid-rapidity region) such that 
\be
t_I = t_i +\gamma_\psi \tau_F~;~{\vec{x}_\psi}^I = ({\vec{r}_\psi}^I,z_\psi^I)
= (r_\psi^I, \phi_\psi^I, z_\psi^I) \nonumber\\
{\vec{v}_\psi}^I =\vec{v}_\psi = ({\vec{v}_\psi}^T,0) = (v_T,0,0) \nonumber\\
F=\vec{v} \cdot \hat{v}_\psi = (z \hat{e}_z/t) \cdot {\hat{v}_\psi}^T =0
\ee
where $\tau_F \approx 0.89~{\rm{fm/c}}$ is the {\it formation} time of the 
bound state
in the $c \bar c$ barycentric frame. Of course, the concept of formation time
was not utilized while drawing the graphs 1-8 in Sec.IV.

\noindent {\ud{\bf Kinematics of $J/\psi$ Trajectory}}:~ The position $3$ vector 
at general instant $t$ becomes 
\be
\vec{x}_\psi \equiv \left[ \vec{r}_\psi, z_\psi \right]
= \left[ {\vec{r}_\psi}^I +(t-t_I) {\vec{v}_\psi}^T, z_\psi^I \right]
\ee
The transverse trajectory will  hit the cylinder of radius $R=R_I$ after a time
interval $t_{RI}$ by covering a distance $d_{RI}$ such that
\be
| {\vec{r}_\psi}^I + t_{RI} {\vec{v}_\psi}^T | =R_I~;~ t_{RI} = d_{RI}/v_T
\nonumber\\
d_{RI} = - r_\psi^I  \cos \phi_\psi^I + \sqrt{R_I^2 -{r_\psi^I}^2 
\sin^2 \phi_\psi^I}
\ee
In the fireball frame the full temporal range of our interest is obviously
\be
t_I \le t \le t_{II}~;~ t_{II} ={\rm{min}} (t_I+t_{RI},t_{\rm{life}})
\ee
This prescription was also utilized in~\cite{rev1} and it improves the work 
of~\cite{xu} by incorporating the formation time.

\noindent {\ud{\bf Formulation of $S(p_T)$}}:~
Let us return to the modified dissociation rate derived in Sec.IV. The value
$F=0$ (cf.(46)) greatly simplifies the kinematics of (30). Furthermore, since 
the time-space point $x$ in (21) is to be chosen on the $J/\psi$ 
{\em trajectory} itself hence the notation $\langle \tilde{\Gamma}(x) 
\rangle$ of (34) is equivalent to
\be
\langle \tilde{\Gamma}[t] \rangle \equiv \langle \tilde{\Gamma}(t, p_{{}_T}, 
z_\psi^I) \rangle
\ee
This depends parametrically  on $z_\psi^I$ in view of the longitudinal
flow profile (44), and the upper limit $t_{II}$ ({\it{cf}}. (49))
of the time variable depends on the production configuration
$r_\psi^I$, $\phi_\psi^I$. Then by the law of radioactive decay without 
recombination the effective
survival chance of a chosen $\psi$ meson will be given by the exponential
$e^{-W}$ with
\be
W = \int_{t_I}^{t_{II}} dt~n_g[t]~\langle \tilde{\Gamma}[t] \rangle
\ee
Upon averaging $e^{-W}$ over the production configuration of the $\psi$'s
we arrive at the final expression for the net survival probability
\be
S(p_T) &=& \int_{V_I} d^3 x_\psi^I (R_I^2 -{r_\psi^I}^2 ) e^{-W}/
\int_{V_I} d^3 x_\psi^I (R_I^2 -{r_\psi^I}^2 ) \nonumber\\
d^3 x_\psi^I &=& d r_\psi^I~r_\psi^I~d \phi_\psi^I~d z_\psi^I
\ee

\noindent {\ud{\bf Remarks}}:~
This is a new result showing that due to the inhomogeneity introduced
by the ensuing longitudinal expansion the spatial integration in (52) must
extend over the {\it volume} $V_I = V_i t_I/t_i$ available at the instant of 
creation. Hence the entry concerning the {\it initial} length $L_i$ and radius
$R_i$ of the plasma
becomes essential in Table 2. However, in the {\it absence} of collective flow
the integral needs to run only over the cross sectional {\it area} $A_I$ of the
fireball as was done in Refs~\cite{rev1,xu}.
%%%%%%%%%%%% begin tab,1%%%%%%%%%%%%%%%%%%%%%%%%%%%%%%%%%%%%%%%
\begin{table}[h]
\caption{Colliding nuclei, collision energy, initial length of 
cylindrical QGP, and its radius at RHIC(1) and LHC(1). The length is
assumed to lie in the range $0.1 \le L_i \le 1$ fm since the sea-quarks
of the nucleon are spread over a distance of order $\Lambda \approx 1$ fm.
[We do not use the $L_i$ corresponding to directly Lorentz-contracted, 
disc-shaped nuclei to avoid too large or too small values.]}

%\vskip 0.2in
\begin{tabular}{lcccc}
\hline
 & Nuclei & Energy($\sqrt{s}$)        & $L_i$ & $R_i$ \\
 &        & (GeV/nucleon) & (fm)  & (fm)\\
\hline
RHIC(1)  & ${Au}^{197}$ & 200& $0.1\le L_i \le 1$ &6.98 \\
  &  & & & \\
LHC(1) & ${Pb}^{208}$ & 5000& $0.1\le L_i \le 1$ & 7.01\\
  &  & & & \\
\end{tabular}
\end{table}

\noindent {\ud{\bf Numerical Illustration}}:~ As pointed out in
Ref.~\cite{rev1} the precise value of the $\psi$ meson formation time
$\tau_{{}_F}$ is ambiguous; however we adopt $\tau_{{}_F} \simeq 0.89$ fm/c
as a fair representative. For chosen creation configuration 
of the $\psi$ meson the function $W$ was first computed from (51) and then 
the survival
probability was numerically evaluated using (52). Figs 9 and 10 show the
dependence of $S(p_T)$ on the transverse momentum corresponding to the
LHC(1) and RHIC(1) initial conditions, respectively (cf. Tables 1, 2). The 
dotted $(L_i=0.1$ fm) and dashed ($L_i=1$ fm) curves are computed in the
{\em presence} of longitudinal flow while the solid curve is borrowed from
~\cite{rev1} in the {\em absence} of hydrodynamic flow profile. The main 
features of these graphs may be explained as follows.

\nd \ud{\bf Interpretation} i) In the absence of flow, i.e., for $v=0$, the
$S(p_T)$ solid lines in Figs.9, 10 rise steadily with $p_T$. This is
because $\langle \tilde{\Gamma} \rangle$ of Fig.4, and hence the $W$ integral
of (51), diminish with $p_T$. ii) In the presence of flow with small
initial length $L_i \sim 0.1$ fm the $S(p_T)$ dotted curves in Figs.9, 10
are slightly above the solid line. To explain this we note that the flow
profile $v=z/t$ remains nonrelativistic over short lengths and the destructive
interference between $I_0$ and $I_1$ terms in (36) becomes slightly stronger
(compared to the $v=0$ case). The consequent drop of $\langle \tilde{\Gamma}
\rangle$ or $W$ pushes $S(p_T)$ upwards. iii) In the presence of flow with
large initial length $L_i = 1$ fm the $S(p_T)$ dashed curve lies below
the solid line in the LHC case (cf. Fig.9) but lies above the
solid line in RHIC case (cf. Fig. 10). 
iv) To explain this peculiar contrast between Figs. 9 and 10 we note that
as the expanding cylinder becomes much bigger than $1$ fm the flow profile
$v=z/t$ may become relativistic over its substantial length so that the graphs
of Fig.8 can be employed. At higher temperatures
$(0.8<T<0.6$, say) at LHC the $\langle \tilde{\Gamma} \rangle$ curves are
known to drop with increasing $p_T$. But at lower temperatures
$(0.6<T<.2)$, say) at RHIC the $\langle \tilde{\Gamma} \rangle$ curves rise 
with $p_T$ in Fig.8 whereby a contrast occurs. v) Of course, the relative shift
in $S(p_T)$ due to longitudinal flow is only about 
10\% to 15\% even at $p_T=10$ GeV.
%%%%%%%%%%%%%%%%%%% begin of figure 9%%%%%%%%%%%%%%%%%%%%%%%%%
\begin{figure}[h]
\psfig{file=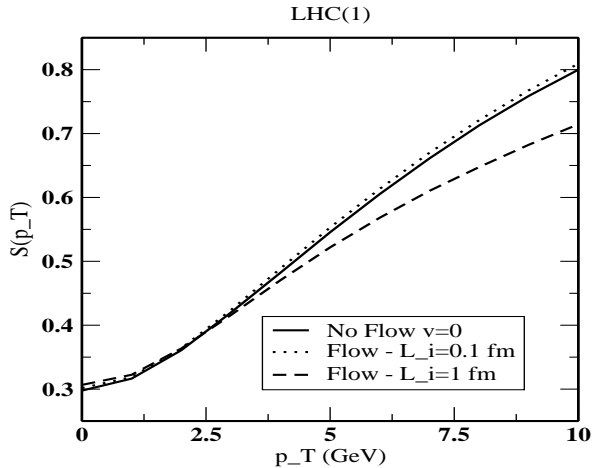,angle=0,height=12cm,width=12cm}
\vskip 0.1in
\caption{The survival probability of $J/\psi$ in an equilibrating
     parton plasma at LHC energy with initial conditions given in 
Table 1. The solid curve is the result of Ref.~\protect \cite{rev1}, 
i.e., in the {\em absence of flow}
while the dotted and dashed curves represent the survival
of $J/\psi$ when the plasma is undergoing longitudinal expansion
with the initial values of the length of the cylinder $L_i=0.1$ fm and
$1$ fm, respectively.}
\end{figure}
%%%%%%%%%%%%%%%%%%% end of figure 9%%%%%%%%%%%%%%%%%%%%%%%%%
%%%%%%%%%%%%%%%%%%% begin of figure 10%%%%%%%%%%%%%%%%%%%%%%%%%
\begin{figure}[h]
\psfig{file=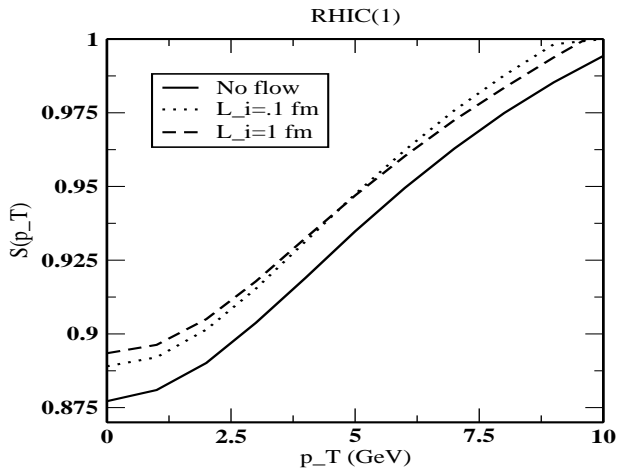,angle=0,height=12cm,width=12cm}
\vskip 0.1in
\caption{Same as Fig.9 but computed for
RHIC initial conditions.}
\end{figure}
%%%%%%%%%%%%%%%%%%% end of figure 10%%%%%%%%%%%%%%%%%%%%%%%%%

\section*{VI Conclusions}  

(a) In this paper we have extended our earlier work~\cite{rev1} by
explicitly including the effect of hydrodynamic expansion profile
on the gluonic dissociation of $J/\psi$'s created in an
equilibrating QGP. The treatment of Secs.III and IV is very general
in the sense that both the fluid velocity $\vec{v}$ and $\psi$
momentum $\vec{p}_\psi$ are arbitrary. Only in Sec.V we specialize to
longitudinal fluid expansion and transverse motion of $\psi$.

(b) Our theoretical formulae for the gluon number density $n_g$(cf.11),
modified $g-\psi$ breakup rate $\langle \tilde{\Gamma} \rangle$
(cf.34), and $\psi$ survival probability $S$ (cf.52) are new. These
are derived by making careful Lorentz transformations between the 
rest frames of the fireball, plasma, and $\psi$ meson.

(c) At specified fugacity $\lambda_g$ the effect of the flow velocity
$\vec{v}$ is to increase the number density of hard gluons (which are
primarily responsible for breaking the $J/\psi$'s) as shown in the
numerical estimate (12).

(d) Our expression (32, 34) of $\langle \tilde{\Gamma} (x) \rangle$ contains
partial wave contributions called $I_0$ and $I_1$ whose mutual interference
is controlled by the anisotropic $\cos \theta_{\psi w}$ factor. This
significantly affects the variation of $\langle \tilde{\Gamma} \rangle$
with $T$, $p_T$ and $\vec{v}$ as depicted in Figs. 3-8.

(e) The presence of longitudinal expansion 
pushes up the graph of $S(p_T)$ compared to the $v \equiv 0$ case 
for nonrelativistic flow appropriate to small initial length
$L_i=0.1$fm. 

(f) However, in the case of relativistic longitudinal flow appropriate
to larger initial length $L_i=1$fm the downward shift of $S(p_T)$
graph at LHC is in sharp contrast to the upward shift at RHIC. Such a
contrast between the behaviours at LHC and RHIC is caused by the 
different initial temperatures generated therein. The relative
shift is, however, is only 10\% - 15\% even at $p_T=10$ GeV.

(g) In a future communication we plan to study the detailed effect of 
transverse expansion of the medium on $S(p_T)$. It is expected that
there will be a more rapid cooling with time and also a possible
constructive interference in (35).

\section*{ACKNOWLEDGEMENTS}  
VJM thanks the UGC, Government of India, New Delhi for financial support.
We thank Dr. Dinesh Kumar Srivastava for useful discussions during 
this work.

\newpage
\section*{Appendix}
\renewcommand{\theequation}{A \arabic{equation}}
% redefine the command that creates the equation no.
\setcounter{equation}{0}  % reset counter
\begin{center}
{\large \bf {EVALUATION OF THE INTEGRAL (28)}}
\end{center}

Our derivation proceeds through the following steps :

\noindent {\ud{\bf First Step}}~We recall the symbols (30, 31), work in 
the $\psi$ meson rest frame, and rewrite (29) as 
\be
\langle \Gamma(x) \rangle = \frac{2}{\pi^3 n_g}
\int_1^\infty \epsilon_\psi^3 dQ^0 {Q^0}^2 
\int_0^{4\pi} d \Omega_{q \psi}~\gamma_\psi
\left[  \frac{}{} 1 + |\vec{v}_\psi| \cos \theta_{q \psi} \frac{}{}
\right] \nonumber\\
\times \sigma_{\rm{Rest}} \sum_{n=1}^\infty \lambda_g^n 
e^{-C_nQ^0+\rho_n \cos \theta_{q w}}
\ee

\noindent {\ud{\bf Second step}}~Next, a partial wave expansion is done of the exponential
\be
e^{\rho_n \cos \theta_{q w}} = \sum_{l=0}^\infty i^l j_l (-i \rho_n)
(2l+1) P_l (\cos \theta_{qw}) \nonumber\\
= \sum_{l=0}^\infty I_l(\rho_n) 4 \pi \sum_{m=-l}^l 
Y_l^{m^\ast} (\Omega_{q \psi}) 
Y_l^m (\Omega_{\psi w}) 
\ee
Here $\rho_n =D_n Q^0$ as before, $j_l$ denotes spherical Bessel functions,
$I_l(\rho_n) = i^l j_l (-i \rho_n)$, the addition theorem has been used for the
Legendre polynomial $P_l (\cos \theta_{q w})$, and $\Omega_{\psi w}$ are the 
polar angles of $\hat{v}_\psi$ with respect to the local flow direction
$\hat{w}$.

\noindent {\ud{\bf Third step}}~ Next, the relevant angular integral 
appearing in (A1,2) reads
\be
\int_0^{4 \pi} d \Omega_{q \psi}~\left[ \frac{}{} 1 + |\vec{v}_\psi| \cos 
\theta_{q \psi} \frac{}{} \right] Y_l^{m^\ast} (\Omega_{q \psi}) \nonumber\\
= \sqrt{4 \pi} \left[ \delta_{l0} + \sqrt{\frac{1}{3}} |\vec{v}_\psi| 
\delta_{l1} \right] \delta_{m0}
\ee
due to the orthogonality of spherical harmonics.

\noindent {\ud{\bf Fourth step}}~Finally, inserting the informations 
(A2,3) back into the starting expression (A1) we obtain 
\be
\langle \Gamma(x) \rangle = \frac{2}{\pi^3 n_g} \int_1^\infty \epsilon_\psi^3
dQ^0 {Q^0}^2 \gamma_\psi~\sigma_{\rm{Rest}} \sum_{n=1}^\infty \lambda_g^n
e^{-c_n Q^0} \nonumber\\
 4 \pi \sum_{l=0}^\infty I_l(\rho_n) \left[ \delta_{l0} + \sqrt{\frac{1}{3}}
|\vec{v}_\psi| \delta_{l1} \right] \sqrt{4\pi} Y_l^0 (\Omega_{\psi w})
\ee
which indeed coincides with (32) of the text.

\end{document}